\documentclass{ws-procs9x6-cpt16}
\begin{document}

\newcommand{\refeq}[1]{(\ref{#1})}
\def\etal {{\it et al.}}
%any other macros go here 

\title{New Searches for CPT Violation in Neutral-Meson Oscillations}

\author{\'Agnes Roberts}
\address{Indiana University Center for Spacetime Symmetries\\
Bloomington, IN 47405, USA}

\begin{abstract}
Classic tests of CPT symmetry have been done in the neutral-meson system. These have been continued with new results in the framework of the Standard-Model Extension, where the coefficient for CPT violation is both boost and direction dependent. Recent contributions and new phenomenological improvements are discussed with possible new contributions for $b$ and $d$ specific correlated decays at the Belle II asymmetric energy collider.
\end{abstract}

\bodymatter

\section{Introduction}
%\phantom{}\vskip10pt\noindent
Neutral-meson interferometry allows precise measurements of possible mass and decay rate asymmetries between particles and antiparticles, thereby testing CPT symmetry. These asymmetries could be due to spontaneous violation of CPT, in which shifts of the rest-mass energy are caused by couplings to a CPT-violating background, arising as a nonzero vacuum expectation value from spontaneous symmetry breaking involving tensor fields.\cite{KosPot} The Standard-Model Extension (SME) is a general quantum field theory framework including Lorentz violating terms in all sectors of the Standard Model. CPT violation implies the violation of Lorentz symmetry.\cite{greenberg} Inspection of the SME terms identifies a single flavor-dependent coefficient $a_{\mu}$ that can be studied with neutral mesons. The flavor-specific relative value for the two valence quarks $\Delta a_{\mu}^{s,b,d}$ provides a unique testing ground for quark sector studies of the SME for $K$, $B$, $D$ neutral meson pairs.\cite{KosPot,neutmesbasics98} These coefficients for CPT violation in neutral mesons have to be direction- and boost-dependent to be consistent with quantum field theory.\cite{neutmesbasics98,neutmesbasics99}

\section{New neutral-meson experiments}

In the last five years, impressive new results have been published on the relevant SME coefficient including its components, taking into account boost and direction dependence. KLOE provided an analysis including all possible spatial binning as well as sidereal binning, and published tight constraints for kaons on all four components of the SME coefficient involving $d$ and $s$ quarks.\cite{KLOE2014} Recently D$\O$ gave improved bounds on the coefficient from like-sign dimuon decays, involving $d$, $s$, and $b$ quarks.\cite{D02015,CPTBsyst} LHCb reported detailed studies and set the best constraints in the $B$ system, also including full component analysis.\cite{LHCb2016} Babar gave new improved results reanalyzing data collected during its operation.\cite{Schubert2016}. Belle II is currently starting operations, targeting at 40 times higher luminosity than the original Belle, with improved pixel and strip detectors providing proper vertex detection even for lower boost studies.\cite{FEI,beltecrep} Lower boost searches open up possibilities for spatial and kinematics studies as well as for the investigation of quantum coherence in the presence of CPT violation. CPT searches are further supported by improvements in data acquisition and analysis. Full event interpretation can allow improved kinematic studies and better neutral particle identification, provide reconstruction of the initial $\Upsilon(4S)$ state, and give better tagging efficiency, giving detailed kinematics information important for momentum-dependent analysis. Belle II carries other advantages compared to the high boost, precision studies at LHCb. It can make specific contributions with high precision studies for the $d$ and $b$ quarks analogous to the detailed studies of correlated decays by KLOE.

%\section{Phenomenology of $B$ factories}
 
In phenomenological studies, more detailed investigations are needed to determine the possibilities for experiments at Belle II. In 2001 Kosteleck\'y introduced phenomenological parameters suitable for any size CPT violation that are phase independent and make explicit the different mechanism by which CPT violation affects the oscillations.\cite{neutmesbasics01} The corresponding parameter is determined in terms of the SME coefficient. This phenomenology included the direction and boost dependence of the relevant CPT-violation coefficient and forms the basis for more appropriate tests. A classical analogue model was also presented for understanding the symmetry violations.\cite{Agnes} 

\begin{figure}
\begin{center}
\includegraphics[width=2in]{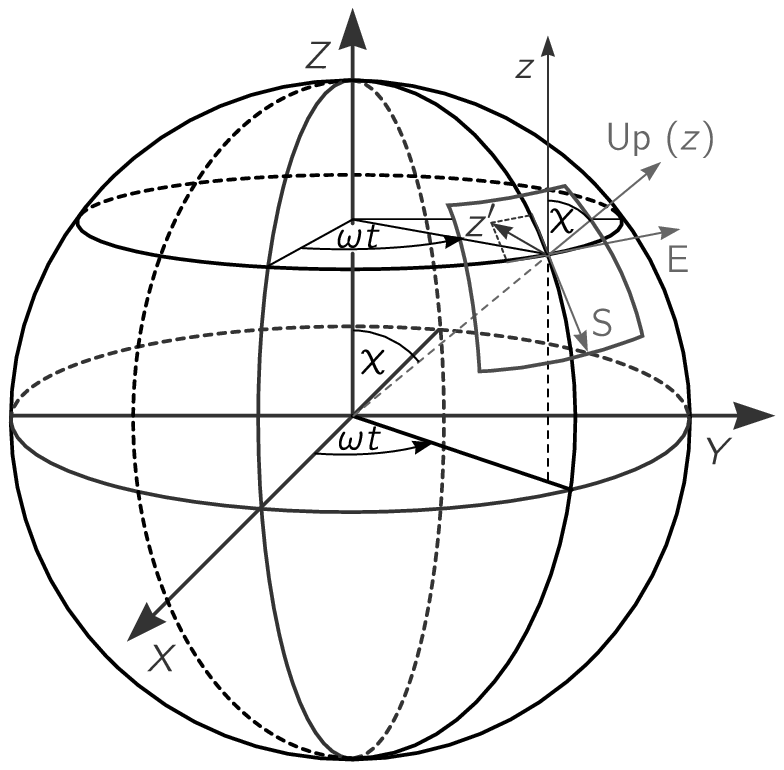}
\includegraphics[width=1.5in]{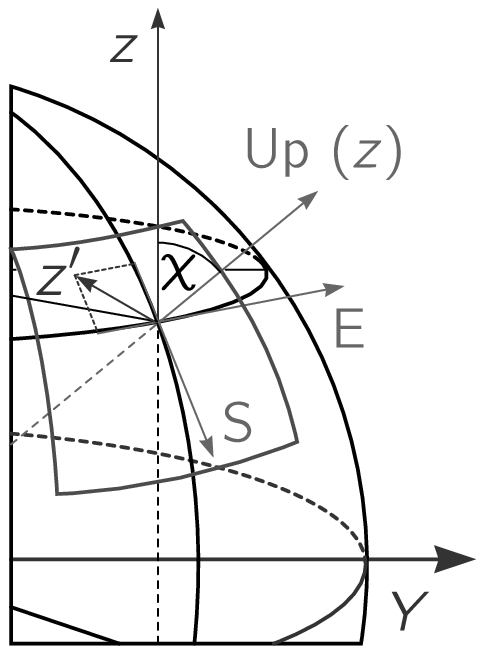}
\end{center}
\caption{Coordinate systems for analyzing neutral-meson SME coefficients.}
\end{figure} 

The 2001 study addressed the correlated mesons separately and is being adapted to give new searches a unified approach. It is extended to investigate new physics in the presence of CPT violation for $B$ factories, where meson antimeson pairs are produced in correlated states and have different oscillation characteristics and experimental issues from uncorrelated mesons and for mesons assumed to experience only CP violation, including the correlated time evolution before the decay of the first meson because of the direction dependent interaction with the nonzero background. The analysis addresses further decay modes, as well as those specific to experiments at Belle II. Details of the geographic location and beam orientation must be considered (see Fig.\ 1),
and there is a difference in how these coordinates are defined at each experiment. It is also necessary to establish the connection between the parameters and definitions traditionally used by the various detectors and those adopted in the SME framework. Since data collection is just beginning, it will be beneficial to initiate sidereal binning of the data and find the optimal binning strategy in concordance with the data analysis.

\end{document}